\documentclass[a4paper,11pt]{article}
\pdfoutput=1 
     \newcommand{\ben}{\begin{equation}} \newcommand{\een}{\end{equation}}
    
\usepackage{jheppub,amsmath,amssymb,bm}
\usepackage{ucs}

\usepackage[utf8x]{inputenc}
\usepackage{textcomp}
\usepackage[T1]{fontenc}
\usepackage{color,accents,stmaryrd}
\usepackage{mathrsfs}
\usepackage{amssymb}
\usepackage{amsmath}
\usepackage{dsfont}
\usepackage{indentfirst}
\usepackage{latexsym}

\newcommand{\cl}{C \kern -0.1em \ell}

\newcommand{\be}{\begin{equation}}
 \newcommand{\ee}{\end{equation}}
 \newcommand{\bea}{\begin{eqnarray}}
 \newcommand{\eea}{\end{eqnarray}}
 
 \newcommand{\nn}{\nonumber}

 \newcommand{\pd}{\partial}

\def\0{\bm0}

 \newcommand{\one}{{\bf 1}}

 \newcommand{\bpsi}{\bar\psi}
 
\long\def\symbolfootnote[#1]#2{\begingroup%
\def\thefootnote{\fnsymbol{footnote}}\footnote[#1]{#2}\endgroup}

\newcommand{\g}{\gamma}

\newcommand{\beq}{\begin{eqnarray}}
\newcommand{\eeq}{\end{eqnarray}}

\definecolor{clearblue}{rgb}{0,0.5,0.9}
\definecolor{orange}{rgb}{1,0.5,0}

\title{\boldmath New Spinor Fields on Lorentzian 7-Manifolds}

\author[a]{L. Bonora}
\author[b]{Rold\~ao da Rocha}

\affiliation[a]{International School for Advanced Studies (SISSA), Via Bonomea
265,
34136 Trieste, Italy}
\affiliation[b]{CMCC, Universidade
Federal do ABC 09210-580, Santo Andr\'e, SP, Brazil}
\emailAdd{bonora@sissa.it}
\emailAdd{roldao.rocha@ufabc.edu.br}
\abstract{ This paper deals with the classification of spinor fields according
to the bilinear covariants in 7 dimensions. 
{}{The previously investigated Riemannian case is characterized by}
{}{
either 
one spinor field class, in the real case of Majorana spinors, or  three
non-trivial classes in the most general complex case. }
In this paper we show that 
by imposing appropriate conditions on spinor fields in 7d manifolds with
Lorentzian metric, the formerly obtained obstructions for new classes of spinor
fields can be circumvented.  New spinor fields classes are then explicitly
constructed. In particular, on 7-manifolds with asymptotically flat black hole
background, {}{these spinors can define a generalized current}
density which further defines a time Killing vector at the spatial
infinity.}
\begin{document}

\maketitle
\flushbottom

\section{Introduction}

Classical spinor fields are characterized by their symmetry properties with
respect to the rotation (Euclidean spacetime) or pseudorotation
(pseudo-Euclidean spacetime) group.
For instance in a 4d Minkowski geometry elementary spinors encompass two
irreducible representations
of the Lorentz group, the Weyl and Majorana spinors, and a reducible one, the
Dirac spinors. With these spinors
we can form bilinears, which in turn fully characterize the spinors themselves
and satisfy the Fierz
identities. If, on the other hand, we reverse the argument and assume that the
defining properties of the spinors are the Fierz identities we find new
surprising and (until recently) unexplored possibilities. The present paper
is in the framework of this new field of research.

Fierz identities were  used by Lounesto to classify spinor fields in
Minkowski spacetime according to the bilinear covariants  in  six disjoint
classes, that encompass all possible spinor fields in {}{4d}  Minkowski spacetime. All
the  spinor classes have been
lately thoroughly characterized \cite{Cavalcanti:2014wia}. The first
three classes  of spinor fields in such classification are referred to as
regular spinor
fields. Their scalar and pseudo-scalar bilinear covariants are different from
zero. The other three classes of
singular spinors are called flag-dipole, flagpole and dipole spinor fields. In
spite of including  Weyl and  Majorana  spinors as very particular cases of
dipole and flagpole spinors respectively  \cite{daSilva:2012wp}, these new
classes further contain genuinely new spinor fields with peculiar dynamics. For
instance other flagpole spinor fields {in these classes}  are eigenspinors of the charge conjugation
operator with dual helicity and may be prime candidates for dark matter 
\cite{exotic,lee2}. Moreover, flag-dipole spinors were found to be  solutions of
the Dirac equation in Einstein-Sciama-Kibble (ESK) gravities \cite{esk}.   A complete overview
of this classification with further applications in field theory and
gravitation can be found in \cite{daSilva:2012wp}. {} {This matter} has been further
explored in the context of black hole thermodynamics, where 
tunnelling methods were studied for {}{ the eigenspinors of the charge conjugation operator having dual helicity \cite{Ahluwalia:2008xi,Ahluwalia:2004ab}, as special type of
flagpoles \cite{bht}}. Experimental signatures of  type-5 spinors in
Lounesto's classification {}{may be} related to the Higgs field at LHC \cite{m1}. 

Motivated by the {}{ new possibilities regarding such recently found new spinors,}
higher dimensional analogues have been recently introduced \cite{bonora,1}, based upon the Fierz identities.
{}{In 7d there are plenty of examples in which new types of spinors
may play a role and become relevant. Thus} {}{Fierz identities may provide an effective framework to attack problems in supergravity and string theory \cite{1},\cite{SUGRA,SUSYBr}, in particular for what concerns compactifications of 11d SUGRA or M-theory to 7 or 4 dimensions.} 
For a comprehensive review on {}{the} physical features related to the
compactification procedure on $S^7$ see, e. g., \cite{Duff:1986hr}. {}{Much in the same way} {}{as in the last decade new physical possibilities, beyond the standard Dirac, Majorana and Weyl spinors, have been introduced and studied in Minkowski 4d spacetime  (see, e. g., Ref. \cite{daSilva:2012wp} for a brief review),}  we aim here to establish the same bottom-up approach, providing the characterization of new classes of spinor fields on Lorentzian 7-manifolds. 

In a previous paper we have constructed a classification of spinors in Riemannian 
7-manifolds \cite{bonora}, 
based upon the fact that only some bilinears are different from zero \cite{1}.
The aim of the present paper is to show
that, when an arbitrary spinor field in Lorentzian 7-manifolds is annihilated by
a linear combination  of the energy operator and the volume element, it singles out a
new class of regular spinors.  

{} {It is remarkable that these spinors can be realized}
as soliton-like solutions in a specific black hole background and to  identify
the current density $J^\mu = \bar\psi\gamma^\mu\psi$ with the Killing vector at
the black hole horizon \cite{mei10}. This is inspired by Kerr and Myers-Perry 5d black holes, which constitutes an appropriate
background wherein a current density interpolates between the time-like Killing vector field 
at the spatial infinity and the null Killing vector fielf on the black hole
event horizon \cite{mei10}. This current density can be realized as a spinor
fluid flow. Here we consider the extension of this construction
to 7d black holes. In fact, by imposing suitable conditions we show that the
current-generating spinors, in a 7d black hole background in a Lorentzian
7-manifolds, precisely realises new classes of spinors, previously precluded by the Fierz
identities on Euclidean 7-manifolds. In particular, the spinor class that we are
going to study in Lorentzian signature has all the associated
bilinear covariants different from zero. Hence these are the 7d analogues of
regular spinors.

{}{Finally, these regular spinors will be shown to further collapse into a specific class of singular ones, when the new regular spinor components satisfy specific constraints}.

{}{The classification we obtain in this paper is in no way exhaustive for 7d Lorentzian manifolds.
Its main aim was to point out that under extremely simple algebraic conditions one can sort out a large class of regular spinors and construct them}.

The paper is organized as follows: in section 2 the classification of spinor
fields in Minkowski spacetime, 
according  to the Lounesto's classification prescription, in 4d is reviewed.
{}{We also summarize the analogous one concerning Euclidean 7-manifolds, wherein complex spinors can be classified in three non-trivial classes, while the classification for real
spin bundles encompasses Majorana spinor fields alone.}

In section 3
the obstruction for the construction of other spinor fields on 7-manifolds is shown
to be circumvented when a different spacetime signature is taken into account,
by imposing certain conditions on the spinor fields. 
Subsequently we show that these new spinor fields can be explicitly constructed
as soliton-like solutions in the framework of a 7d black-hole background. They
are implicitly defined by a current of probability (which further defines
the time-like Killing vector at the spatial infinity) and explicitly constructed {}{via the reconstruction theorem}. 
{}{Comparing with the case of Riemannian 7-manifolds,} we see that the number and types of
spinor field classes, classified by the bilinear covariants via the Fierz
identities, are signature dependent. Section 4 is devoted to our
concluding remarks and outlooks.

\section{General Bilinear Covariants and Spinor Field Classes in 7d}

An oriented manifold $(M,g)$ 
 and its tangent bundle $
TM$ admits an exterior bundle 
$\bigwedge(TM)$. The Clifford product involving an arbitrary 1-form field $ v \in
\sec\bigwedge^1(TM)$ and an arbitrary form   
$a \in \sec\bigwedge(TM)$ is specified by a combination of the exterior product and the contraction, namely, $ v \circ a = v \wedge a+ v 
\lrcorner\, a $. By taking the particular case of  Minkowski spacetime, the
basis
 $\{ e^{\mu }\}$ represents a section of the coframe 
bundle
${P}_{\mathrm{SO}_{1,3}^{e}}(M)$. 
 Classical {Dirac} spinor fields are elements that carry the 
$\rho = {(1/2,0)}\oplus {(0,1/2)}$ representation of the Lorentz group. For any 
spinor field 
$\psi \in \sec {P}_{\mathrm{Spin}_{1,3}^{e}}(M)\times_{\rho}
\mathbb{C}^{4}$, the bilinear covariants  are given by:
\begin{subequations}
\begin{eqnarray}
\sigma &=& \bar{\psi}\psi\,,\label{sigma}\\
J_{\mu } e^{\mu }=\mathbf{J}&=&\bar{\psi}\gamma _{\mu }\psi\, e
^{\mu}\,,\label{J}\\
S_{\mu \nu } e^{\mu}\wedge e^{ \nu }=\mathbf{S}&=&\tfrac{1}{2}i\bar{\psi}\gamma
_{\mu
\nu }\psi \, e^{\mu }\wedge  e^{\nu }\,,\label{S}\\
 K_{\mu } e^{\mu }=\mathbf{K}&=&i\bar{\psi}\gamma_{5}\gamma _{\mu }\psi
\, e^{\mu }\,,\label{K}\\\omega&=&-\bar{\psi}\gamma_{5}\psi\,,  \label{fierz}
\end{eqnarray}\end{subequations}
where $\bar\psi=\psi^\dagger\gamma_0$,
$\gamma_5:=\gamma_0\gamma_1\gamma_2\gamma_3$ and $\gamma_{\mu }\gamma _{\nu
}+\gamma _{\nu }\gamma_{\mu }=2\eta_{\mu \nu }\mathbf{1}$. 
 The Fierz identities read
\begin{equation}\label{fifi}
\mathbf{K}\wedge\mathbf{J}=(\omega+\sigma\gamma_{5})\mathbf{S},\qquad\mathbf{J}^
{2}=\omega^{2}+\sigma^{2},\qquad\mathbf{K}^{2}+\mathbf{J}^{2}
=0=\mathbf{J}\cdot\mathbf{K}
\,.  
\end{equation}
\noindent When either $\omega\neq0$ or $\sigma\neq0$ [$\omega=0=\sigma$] the
spinor field $\psi$ is named regular [singular] spinor. 

Lounesto classified spinor fields into six disjoint classes. In the
classes (1), (2), and (3) beneath it is implicit that $\mathbf{J},$ $\mathbf{K}$
and $\mathbf{S}$ are simultaneously different from zero, and in the classes (4),
(5), and (6) just $\mathbf{J}\neq 0$:
\begin{itemize}
\item[1)] $\sigma\neq0,\;\;\;\omega\neq0$\qquad\qquad\qquad\qquad\qquad4) $
\mathbf{K}\neq 0,\;\;\;\mathbf{S}\neq0$\;\;\; $(\sigma=\omega=0)$%
\label{Majorana 11}
\item[2)] $\sigma\neq0,\;\;\;
\omega = 0$\label{dirac1}\qquad\qquad\qquad\qquad\qquad5) $\mathbf{K}=0,\;\;\;
\mathbf{S}\neq0$\;\;\;$(\sigma=\omega=0)$
\label{tipo41}
\item[3)] $\sigma= 0, \;\;\;\omega \neq0$\label{dirac21}
\qquad\qquad\qquad\qquad\qquad\!6) $\mathbf{S}=0,
\;\;\; \mathbf{K} \neq 0$\;\;\;$(\sigma=\omega=0)$
\end{itemize}
\noindent  
Singular spinor fields of types-4, -5, and -6 are flag-dipoles, flagpoles and
dipole
spinor fields, respectively.  {}{It is worth to emphasize that in classes (4), (5) and
(6) 
the vectors $\{{\bf J}, { \bf K}\}$
can not be elements of a basis for Minkowski spacetime and collapse into a null-line.  By defining {\bf J} as the pole, flagpoles are hence defined in the class-(5), since for this case ${\bf K} = 0$ and ${\bf S}\neq 0$. The 2-form  ${\bf S}$ is interpreted as a plaquette, namely, a flag-pole. In the case of the type-4 spinor fields, 
both ${\bf S}$ \emph{and} ${\bf K}$ are not equal zero, and they form a flag-dipole. Both concepts encompass Penrose flagpoles \cite{daRocha:2007pz}.} The first physical example of flag-dipole spinor
fields has been recently found to be a solution of the Dirac equation in ESK
gravities \cite{esk}. Moreover, Majorana and Elko spinor fields reside in the
class of type-5 spinors, whereas Weyl spinor fields are a particular example of a type-6
dipole spinor fields, that further encompass pure spinors as well
\cite{daSilva:2012wp}. {}{The characterization of new singular spinor fields in 
Lounesto's classes has introduced new fermions, including mass dimension one matter fields, that have been studied in}
\cite{lee2,exotic,esk,Cavalcanti:2014wia,daSilva:2012wp,bht}.  The  most general types of
spinor fields in each class of
Lounesto's classification have been developed in \cite{Cavalcanti:2014wia}. 
For singular spinors the Fierz identities (\ref{fifi}) read  \cite{Cra}:
\begin{align}
\!\!\!\!\!{\rm Z}^{2}   =4\sigma {\rm Z},\quad {\rm Z}\gamma^{\mu}{\rm
Z}=4J^{\mu}{\rm Z},\quad{\rm Z}i\gamma^{5}\gamma^{\mu}{\rm Z}    =4K^{\mu}{\rm
Z},\quad {\rm Z}i\gamma^{\mu\nu
}{\rm Z}=4S^{\mu\nu}{\rm Z},\quad 
 {\rm Z}\gamma^{5}{\rm Z}=-4\omega {\rm Z}\,,
\end{align}
{}{where ${\rm Z}=\sigma + \mathbf{J} +i\mathbf{S}+i\mathbf{K}\gamma_{0123}+\omega \gamma_{0123}$.}

In arbitrary manifolds with $(p,q)$ signature
$p+q=n=\dim M$), given a spin bundle {\it S}\, 
and  ${\gamma^{n+1}}\in {\rm sec}({\rm End}(S))$ \cite{1}, 
spin projectors $\Pi_\pm= \frac{1}{2}( I  \pm {\gamma^{n+1}})$ can be constructed. They provide the
spin bundle splitting  $ 
S=S^+\oplus S^-$, where $
S^\pm= \Pi_\pm (S)$. 
Sections of $S^\pm$ are named
  {Majorana-Weyl  spinors}  when $p-q\equiv 0\mod 8$, whereas sections of $S^+$
are known as Majorana
spinors when  
 $p-q\equiv 7\mod 8$. 

Let an orthonormal coframe be given, in 7d, by $\{e^a\}_{a =0}^{6}$. 
Hereupon we adopt the notation
$\gamma_{\rho_1\rho_2\ldots\rho_k}=\gamma_{\rho_1}\gamma_{\rho_2}\cdots\gamma_{
\rho_k}$ and $e^{\rho_1\ldots \rho_k} = e^{\rho_1}\wedge\cdots\wedge
e^{\rho_k}$.  In general  the spinor conjugation reads $
\bar\psi=\psi^\dagger a^{-1}$, for $a\in
\mathcal{C}\ell_{p,q}^*=\mathcal{C}\ell_{p,q}\backslash{\{0\}}$ 
where $\mathcal{C}\ell_{p,q}$ denotes the Clifford bundle on a spacetime with signature $(p,q)$.  Given
$\psi,\psi'\in \Gamma(S),$ and given a bilinear form $B$ on $\Gamma(S)$,  the most general bilinear  on $S$ read: 
\beq\label{formab}
\beta_k(\psi,\psi^\prime)=B(\psi,\gamma_{\rho_1\dots\rho_k} \psi^\prime)=
{\bar{\psi}}{\gamma}_{\rho_1\dots\rho_k} {\psi^\prime}\,.
\eeq\noindent  

Now, generalized bilinear covariants are defined by  \cite{1,bonora}
 \beq
\label{ddd}
\varphi_k:= \frac{1}{k!}B (\psi,\gamma_{\rho_1\ldots
\rho_k}\psi)e^{\rho_1\ldots \rho_k} = \bar\psi\gamma_{\rho_1}\ldots
\gamma_{\rho_k}\psi\,e^{\rho_1\ldots \rho_k}
\in\sec\bigwedge^k(TM)\,.\eeq\noindent  For $\psi$ a Majorana spinor, the
forms $\varphi_k$ equal zero except when either  $k=0$ or $k=4$ \cite{bonora,1}.
{ I.e. such class of Majorana spinor fields, according to the bilinears in the
Clifford bundle ${\cal C}\ell_{7,0}$, is provided by:
\beq
\!\!\!\!\!\!\!\!\!\!\!\!\!\!\varphi_0\neq 0, \quad \varphi_1=0, \quad\varphi_2=
0, \quad\varphi_3= 0,\quad
\varphi_4\neq 0,\quad\varphi_5=0, \quad\varphi_6=0, \quad
\varphi_7=0\,.\label{class0}\eeq\noindent} 

The bilinear
covariants in Eq.(\ref{ddd}), except $\varphi_0$ and $\varphi_4$, were shown to
be null on Euclidean 7-manifolds in Ref.\cite{bonora}, { as a consequence of}
the geometric Fierz identities \cite{1}. {We will see in the next section that 
these obstructions  can be circumvented in a different signature, once
appropriate conditions are imposed. In other words, we obtain the bilinear
covariants for the associated spinor fields when appropriate conditions are
enforced on them \cite{wei}. More precisely,  by imposing that the spinor field
is annihilated by the one of the operators
$\gamma^0\pm\gamma^8$, we can prove that new distinct classes do exist in the
above classification.} 

{}{It is still worth to mention that 
real representations associated to the Clifford algebra over the (6,1)
Lorentzian space  admit a quaternionic structure induced by globally defined
operators $J_i$ ($i = 1,2,3$) \cite{1}. Such structure can be used to construct
the following bilinear covariants:
\beq
\label{dddj}
\mathring\varphi_k:= \frac{1}{k!}B (\psi, J_i\circ \gamma_{\rho_1\ldots
\rho_k}\psi)e^{\rho_1\ldots \rho_k} := \mathring{\bar\psi}\gamma_{\rho_1}\ldots
\gamma_{\rho_k}\psi\,e^{\rho_1\ldots \rho_k}
\in\sec\bigwedge^k(TM)\,.\eeq\noindent 
Here the product ``$\circ$'' is the standard product in the spin bundle
End($S$).  
In \cite{1}, for Euclidean 7-manifolds, Eq.(\ref{dddj}) was proved to be related
to (\ref{ddd}) by Hodge duality. However it does not necessarily hold
for the case of Lorentzian manifolds. 
In the latter case  the quaternionic structure in (\ref{dddj}) induced by the
$J_i$ can be
taken into account. Hence there is a $S^2$-family of  complex structures that
can define an $S^2$ family of bilinear covariants $\mathring\varphi_k$.
{}{We remark, nevertheless, that} this can be equivalently realized by incorporating the $J_i$ in the conjugate $\bar\psi$ of the spinor $\psi$ in Eq.(\ref{formab}), denoted by
$\mathring{\bar\psi}$ in Eq. (\ref{dddj}), clearly defining
new equivalent spinor conjugates. A thorough discussion concerning quaternionic
structures on manifolds of with different signatures can be found in \cite{1}}.

\section{New Classes of 7d Spinors}


In this section we show that in a Lorentzian 7-manifold some obstructions found
in the Euclidean case for the existence of general spinors can be circumvented.
In fact we show that {}{(analogues of)} regular spinors can be effectively
constructed.

Let us define a {}{vielbein} basis of the Clifford algebra over the (6,1) Lorentzian
space  as follows
\be \gamma^0=i\sigma^1\otimes\mathbb{I}\,,\quad
\gamma^6=\sigma^3\otimes\mathbb{I}\,,\quad
\gamma^a=-\sigma^2\otimes\gamma_\diamond^{a-1}\,,\quad
a=1,\ldots,5\,,\label{sieb}\ee
where $\mathbb{I}$ denotes hereupon  $4\times 4$ identity operator and
$\gamma_\diamond^{\mu}$ are provided by  \cite{wei,budinich}
\be \label{funf}\gamma_\diamond=i\sigma^1\otimes\one_2\,,\quad
\gamma_\diamond^4=\sigma^3\otimes\one_2\,,\quad
\gamma_\diamond^j=-\sigma^2\otimes\sigma^j\,,\quad j=1,2,3\,\ee
 The  {}{vielbein} basis (\ref{sieb}) is constructed from (\ref{funf}) 
by the method described in \cite{budinich}. Exploiting the fact that bilinear
covariants are representation-independent, 
(\ref{sieb}) is chosen to provide a nicer form for the spinor components, see
(\ref{spinorr} - \ref{7dequal}), under the condition  (\ref{condicao}) below.

Now the relevant spinor field is represented as
\be
\label{spinorr} \psi=\left(\alpha_0,\ldots,\alpha_7\right)^\intercal\,\in\sec
{P}_{\mathrm{Spin}_{1,6}^{e}}(M)\times_{\rho}\mathbb{C}^{8}\,,
\ee
where $\rho$ stands for a representation of the associated Lorentz group and the
$\alpha_a$ (${\scriptstyle a \,=\, 0,\ldots,7})$ are complex  functions. The
spinor $\psi$ is required to satisfy the condition 
\bea 
(\gamma^0\pm \gamma^8)\psi=0\,,\label{condicao}
\eea
which yields \cite{mei10}
\bea
\alpha_\mu=\alpha_{\mu+4}, \qquad {\scriptstyle{\mu\, =\, 0, 1, 2, 3}}\,.
\label{7dequal}
\eea {}{This condition is the only linear combination of gamma matrices (or products of gamma matrices)  providing conditions for the spinor components 
that generate new classes of spinor fields.}

By calculating the bilinear covariants in Eqs.(\ref{ddd}) one can
straightforwardly  realize  that all the bilinears are {\it generically}
different from zero 
{}{(unless very particular constraints among the spinor components
hold. The  conditions are derived in the Appendix.)}. In fact, for these kind of
spinors, generically, we have
\beq
\!\!\!\!\!\!\!\!\!\!\!\!\!\!\varphi_0\neq 0, \quad \varphi_1\neq 0,
\quad\varphi_2\neq  0, \quad\varphi_3\neq  0,\quad
\varphi_4\neq 0,\quad\varphi_5\neq 0, \quad\varphi_6\neq 0, \quad
\varphi_7\neq 0\,.\label{class}\eeq\noindent
Hence, spinors (\ref{spinorr}) associated to the above bilinear covariants play
the role of regular spinors on Lorentzian 7-manifolds. \footnote{{}{The conditions (\ref{class})
and the additional ones in Appendix are pointwise, so the question arises of 
such spinors being globally defined. The spacetime spin manifolds of major
interest
are Lorentzian simply connected manifold with trivial holonomy,
that is maximally symmetric spaces (dS, AdS or Minkowski). In such cases the
conditions
(\ref{class}) being true in a single chart is enough (see, for instance,  the
explicit 
example shown below). If the manifold $M$ has a more complicated topology one has
of course to check
that the conditions are preserved by the relevant transition functions.}} 
{}{In Appendix we explicitly calculate the bilinear covariants defined in (\ref{ddd}) when the spinor (\ref{spinorr}) is subject to the condition (\ref{condicao})}.
and we prove that all the bilinear covariants are different from zero
unless very specific constraints are satisfied. Such possible exceptions are: 
\beq
1):&& \text{$\varphi_0=0$, \quad if \quad $\alpha_2\overline\alpha_1 +
\alpha_1\overline\alpha_2+\alpha_4\overline\alpha_3+\alpha_3\overline\alpha_4=0$
}\label{112}
;\\
2):&& \varphi_6=0, \quad\;{\rm if}\;\quad 
\begin{cases}
&\alpha_2\overline\alpha_1 -
\alpha_1\overline\alpha_2+\alpha_4\overline\alpha_3-\alpha_3\overline\alpha_4=0\
,\\
&\left| \alpha_1\right| ^2+\left| \alpha_2\right| ^2\!-\!\left| \alpha_3\right|
^2-\left| \alpha_4\right|
   ^2=0\,\;
\end{cases}\label{113}\\
3):&& \varphi_7 = 0, \quad {\rm if}\quad  \alpha_2\overline\alpha_1 -
\alpha_1\overline\alpha_2+\alpha_4\overline\alpha_3-\alpha_3\overline\alpha_4=0
\,.\label{114}
\eeq
In the particular case where the above conditions 1), 2), and 3) are
simultaneously satisfied, by analysing the terms in (\ref{0-form}),
(\ref{6-form}) and (\ref{7-form}) and equating them to zero we get
\beq
\frac{\left| \alpha_4\right| ^2\left| \alpha_3\right| ^2}{\left| \alpha_3\right|
^2+\left| \alpha_4\right|
   ^2}=\left| \alpha_2\right| ^2\left(1-\left| \alpha_2\right|
^2\right)\,.\label{constraint}\eeq\noindent If this condition is satisfied, then
$\varphi_0 = 0 = \varphi_6 = \varphi_7$.

A clarification is in order at this point. If no condition such as
Eq.(\ref{condicao}) is imposed on a generic spinor $\psi$, all bilinear
covariants  are generically nonvanishing. Indeed, if
no condition is  imposed on the spinor $\psi$, then the spinor components
(\ref{spinorr}) are generic and 
will satisfy (\ref{class}). {}{Condition (\ref{condicao}), however, assures the 
computational feasibility of finding
physical solutions of the Dirac equation in Lorentzian manifolds. Since we  
show that the possible 
spinor classes are restricted to eight, instead of the 128 initial possible
ones, 
Moreover, the spinor components are four linearly independent
ones due to (\ref{7dequal}), instead of the initial eight ones.
{}{All this makes the quest for solutions much more manageable.
It is obvious from the above that our aim in this paper is not to exhaust all
the possibilities for spinor fields in 7d Lorentzian manifolds, but rather to
select a subclass of them with remarkable properties, in particular such that
they can be identified as regular spinors.}

Among these properties let us cite also the difference with  the case of
4d Minkowski spacetime. Here  an arbitrary spinor (\ref{spinorr}) will trivially
satisfy (\ref{class}); however, the spinor (\ref{spinorr}) with components
satisfying (\ref{7dequal}) still satisfy (\ref{class}), a result that has no
analogue in the standard Lounesto's 4d classification \cite{daSilva:2012wp}.
 
The spinor $\psi$ can be explicitly constructed from the above bilinears. In
fact, a spinor can be determined up to a phase from the bilinear covariants, by
the inversion theorem \cite{Cra}. Given $\psi$ let us consider an aggregate
\cite{bonora}
\beq\label{agreg}
 {\rm Z}= \sum_{k=0}^7\;\bar\psi \g_{\rho_1\ldots \rho_k}\psi\;e^{\rho_1\ldots
\rho_k},
\eeq where the sum is ordered in $k$. Starting from an arbitrary 7d spinor
$\tau$ satisfying $\widetilde{\tau^{*}}\psi\neq 0$, the original spinor $\psi$
can be recovered from its aggregate (\ref{agreg}). The spinor $\psi$ 
and the multivector field\footnote{It is worth to mention that $ 
{\rm Z}\tau$ is, a priori, a multivector that we {}{can} prove to be an element of a
minimal left ideal in the associated Clifford algebra. Hence it is in fact an
algebraic spinor, according to the Chevalley construction.} ${\rm Z}\tau$  are
equivalent up to a scalar: 
\begin{equation}
\psi=\frac{1}{2\sqrt{\tau^{\dagger}\gamma_{0}{\rm Z}\tau}}\;e^{-i\vartheta}{\rm
Z}\tau, \label{3}%
\end{equation}
\noindent where 
$e^{-i\vartheta}={2}({\tau^{\dagger}\gamma_{0}{\rm
Z}\tau})^{-1/2}\tau^{\dagger}\gamma_{0}\psi\in {\rm U(1)}$. This generalizes to
7d the well known reconstruction theorem (Takahashi algorithm) \cite{Cra},
leading to the 4d equivalent of (\ref{3}).  {}It is
worth to mention that a comprehensive discussion and a complete proof regarding
the reconstruction theorem for more general cases is {}{carried out} in
\cite{Babalic:2014cea}.

Next we would like to show that such types of new fermions may appear, {}{as soliton-like solutions}, in a suitable black hole background.
The metric for stationary and axisymmetric black holes in 7d can be expressed by
\cite{mei10}
\bea\label{metrica1}  \!\!\!\!\!\!\!\!\!\!\!\!\!\!\!\!\!\!\!\!\!ds^2&=&-f_t(\Pi
dt+f^id\phi_i)^2+f_rdr^2
+f_{1}^Ad\theta_A^2
+g_{11}\Big[d\phi_1\!-\!w_1 dt+g_{12}(d\phi_2-w_2 dt)
\nn\\&&+g_{13}(d\phi_3\!-\!w_3 dt)\Big]^2\!\!+g_{22}\Big[d\phi_2\!-\!w_2
dt+g_{23}(d\phi_3-w_3 dt)\Big]^2
+g_{33}(d\phi_3-w_3 dt)^2\,,
\eea
where $\Pi = 0$ or $1$, $A=1,2$ and the functions $f_a,  
g_{ab}$ and $w_a$ depend only upon the radius $r$ and the latitudinal angles
$\theta^A$. The latter can be always chosen to be positive
definite near the black hole horizon \cite{mei10}.    In such cases $w^a
\rightarrow\Omega^a$ as $r\rightarrow R\,,$ \cite{mei10}  
where $R$ denotes a coordinate singularity and $\Omega^a$ stands for the angular
velocity of the black hole in the
$\phi^a$ direction. 
Hence (\ref{metrica1}) can be rewritten in terms of vielbeins \be
ds^2=\eta_{AB}^{~}e^Ae^B\,,\quad A,B=0,\ldots,6\,,
\label{metric.vielbein}\ee
where $\eta={\rm diag}\,(-+\cdots+)$. In
general, (\ref{metric.vielbein}) is only well defined near the black hole
horizon, where all the
vielbeins are real \cite{mei10}.

Now, using an arbitrary spinor satisfying (\ref{condicao}), let us consider the
vector field  
\be
\xi^\rho =b_\psi \bpsi\gamma^\rho\psi\,.\label{definxi}
\ee
In the particular case where $b_\psi =1$,  this is the current density $J^\rho =
\bpsi\gamma^\rho\psi$. In the most general case,  $b_\psi $ is some scalar. The
spinor $\psi$ is not necessarily a 
regular fermion \cite{bonora}, and in particular it can be 
the higher dimensional version of flag-dipoles, flagpoles or dipoles singular
spinors fields \cite{daSilva:2012wp}.
{Given that the spinor field $\psi$
obeys one of the two conditions (\ref{condicao}), the current density $J^\rho =
\bpsi\gamma^\rho\psi$ reduces to the expression 
\begin{eqnarray}\label{killi}
J^\rho \pd_\rho
&=&{2 \left(\left| \alpha_1\right| ^2+\left| \alpha_2\right| ^2+\left|
\alpha_3\right| ^2+\left| \alpha_4\right|
   ^2\right)}\pd_t\,.
\end{eqnarray} }
\noindent Although this is outside the topic of this paper we notice that, since $w^a$ are
constants on the
horizon \cite{mei10}, {the vector field $\xi=\xi^\rho\partial_\rho$ in Eq.
(\ref{definxi}) -- a multiple of the current density in Eq. (\ref{killi}) --}
can  be identified with  the null Killing vector on the black hole horizon.
In fact the vector field $\xi$ can be shown to  interpolate between the time Killing
vector at the spatial infinity and the null Killing vector on the
horizon \cite{wei}. In addition, $\nabla_\mu\xi^\mu =0$, what justifies calling
$\xi$ a conserved current.

A 1-form current density 
$\xi^\rho  = b_\psi  \bar\psi\gamma^\rho\psi$ has been constructed, being the
conserved current of a particular spinor field. In the background of a
stationary black hole the current density vector
field always approaches the null Killing vector at the  horizon. When
the black hole is asymptotically flat and when the coordinate
system is asymptotically static, the same vector field also
becomes the time Killing vector at spatial infinity \cite{wei}. The
required constraint on the spinor field unblock the obstructions that the Fierz
identities impose on new classes of spinors fields. {In fact we have proved
that, in the context of this paper, the bilinear covariants (\ref{ddd}) are
different from zero. It is worth to emphasize that when {the above mentioned coefficient 
$b_\psi$ is non-vanishing, we can view the left hand side of Eq.(\ref{killi}) as
a well-defined Killing vector}. 

We recall that these results, different from the ones obtained in
\cite{bonora}, are explained by the fact that the metric used here is Lorentzian,
while in \cite{bonora} it was Euclidean.}

\section{Concluding Remarks and Outlook}

 The
geometric Fierz identities \cite{bonora,1} are well known to limit the number of
classes of spinor
fields according to the bilinear covariants.
{}{In a previous paper we have investigated Majorana
spinor fields in Euclidean 7-manifolds, proving that the geometric Fierz
identities forbid the existence of more
than one spinor field class (\ref{class}) in the real case \cite{bonora}, while
three non-trivial classes can exist in the complex case. 
However the obstructions 
that preclude the existence of further spinor field classes in Euclidean
7-manifolds can be attenuated in a different spacetime metric signature, when
conditions are imposed on the spinor fields.}

{}{We have achieved this by imposing one of the conditions (\ref{condicao}), for in this case the associated spinor field have all bilinear covariants non-vanishing. This property
has remarkable implications. In Lounesto's spinors classification in 4d
Minkowski spacetime, the bilinear covariants for spinors of type-1 are all  non zero, and
regular spinors play the role of the standard Dirac spinor, describing, for
instance, the electron in the Dirac theory. Here, in a 7-manifold with
Lorentzian signature,   calculating the bilinear covariants for a spinor under
the condition (\ref{condicao}),  we prove that all bilinear covariants are generically non-vanishing,
We deduce that in a generalized classification  of spinors in Lorentzian 7-manifolds
according to the observables (\ref{ddd}), the spinor
(\ref{spinorr}) with the constraint (\ref{condicao})  can be interpreted as the
analogues of the regular spinor in 4d.}

{}{As we have remarked, an exception is when the constraint (\ref{constraint}) holds. In this case we have a restriction $\varphi_k = 0$, for either $k=0$ or $k=6$ or $k=7$}. However, preliminary calculations
show that some of the conditions (\ref{112},
\ref{113}, and \ref{114}) can be only achieved for spinors that are solutions
of the Dirac equation in Lorentzian 7-manifolds with torsion. Hence, without
torsion (or a Kalb-Ramond background field), our results show that there is only
one class of regular spinors.

It should also be recalled} that we have studied the classification of
spinors related to the standard bilinear covariants based upon (\ref{ddd}). One
could also consider an $S^2$ family of isomorphic classifications provided by
the quaternionic structures $J_i$ in (\ref{dddj}). 
{}{These cases are already included in the previous analysis for we can
always} incorporate the
representations of the $J_i$ in the conjugate spinor in (\ref{dddj}). Since the
bilinear covariants are representation independent, 
{}{the spinor classifications
induced via the quaternionic structures are indeed isomorphic to those obtained above.}

Once the classes of spinor fields on both Euclidean and Lorentzian
7-manifolds have been identified, 
{}{a further question arises by considering
inequivalent spin structures on 7-manifolds}. They are well known to induce an
additional term on the associated Dirac operator, related to a $\check{\rm
C}$ech cohomology class \cite{exotic,alex}.  Although the existence and
classification in the above sections is not modified by considering inequivalent
spin structures, the dynamics associated to spinor fields in each class can
manifest important modifications, which we shall discuss in a forthcoming
publication.

\acknowledgments

R. da Rocha is grateful for the CNPq grants No. 303027/2012-6, No.
451682/2015-7, No. 473326/2013-2, and FAPESP grant No. 2015/10270-0, and to INFN
grant ``Classification of Spinors'', which has provided partial support.

\appendix
\section{Appendix}

When the spinor (\ref{spinorr}) under the condition (\ref{condicao}) is taken
into account, we can explicitly calculate the bilinear covariants defined in
(\ref{ddd})  as:
\beq
\varphi_0&=&2i\left(\alpha_2\overline\alpha_1 +
\alpha_1\overline\alpha_2+\alpha_4\overline\alpha_3+\alpha_3\overline\alpha_4\right)\\&&\label{0-form}\\
\varphi_1&=&\bar\psi\gamma_{k}\psi\;e^{k}\nonumber\\
&=&2 \left(\left| \alpha_1\right| ^2+\left| \alpha_2\right| ^2+\left|
\alpha_3\right| ^2+\left| \alpha_4\right|
   ^2\right)\;e^0-2 \left(\left| \alpha_1\right| ^2+\left| \alpha_2\right|
^2+\left| \alpha_3\right| ^2+\left| \alpha_4\right|
   ^2\right)\;e^5
   \eeq\noindent It implies that unless $\left| \alpha_1\right| ^2+\left|
\alpha_2\right| ^2+\left| \alpha_3\right| ^2+\left| \alpha_4\right|
   ^2=0$, namely, when $\alpha_1 = 0 = \cdots = \alpha_4$, the 1-form bilinear
covariant is always different from zero.

Now, the 2-form bilinear covariant explicitly reads:
\beq
\varphi_2 &=& \bar\psi\gamma_{k_1k_2}\psi\;e^{k_1k_2}\nonumber\\
&=& 2i\left(\alpha_2\overline\alpha_1 +
\alpha_1\overline\alpha_2+\alpha_4\overline\alpha_3+\alpha_3\overline\alpha_4\right)\;e^{05}-2\left(\alpha_2\overline\alpha_1 -
\alpha_1\overline\alpha_2+\alpha_4\overline\alpha_3-\alpha_3\overline\alpha_4\right)\;e^{12}
\nonumber\\
&&-2i\left(\left| \alpha_1\right| ^2-\left| \alpha_2\right| ^2+\left|
\alpha_3\right| ^2-\left| \alpha_4\right|
   ^2\right)\;e^{13}+2\left(\left| \alpha_1\right| ^2-\left| \alpha_2\right|
^2+\left| \alpha_3\right| ^2-\left| \alpha_4\right|
   ^2\right)\;e^{14}\nonumber\\
&&+2i\left(\left| \alpha_1\right| ^2-\left| \alpha_2\right| ^2+\left|
\alpha_3\right| ^2-\left| \alpha_4\right|
   ^2\right)\;e^{16}+2\left(\left| \alpha_1\right| ^2+\left| \alpha_2\right|
^2+\left| \alpha_3\right| ^2+\left| \alpha_4\right|
   ^2\right)\;e^{23}\nonumber\\
&&+2i\left(\left| \alpha_1\right| ^2+\left| \alpha_2\right| ^2+\left|
\alpha_3\right| ^2+\left| \alpha_4\right|
   ^2\right)\;e^{24}-2\left(\left| \alpha_1\right| ^2+\left| \alpha_2\right|
^2+\left| \alpha_3\right| ^2+\left| \alpha_4\right|
   ^2\right)\;e^{26}\nonumber\\
&&+2\left(\alpha_2\overline\alpha_1 +
\alpha_1\overline\alpha_2+\alpha_4\overline\alpha_3+\alpha_3\overline\alpha_4\right)\;e^{34}
+2i\left(\alpha_2\overline\alpha_1 +
\alpha_1\overline\alpha_2+\alpha_4\overline\alpha_3+\alpha_3\overline\alpha_4\right)\;e^{36}\nonumber\\&&
-2\left(\alpha_2\overline\alpha_1 +
\alpha_1\overline\alpha_2+\alpha_4\overline\alpha_3+\alpha_3\overline\alpha_4\right)\;e^{56}
\eeq
Since the above spinor components $\alpha_\mu$ are functions of the point in the
manifold, then all terms of the above 2-form must be zero in order for
$\varphi_2$ to vanish. It occurs if and  only  if all $\alpha_\mu=0$. Therefore
$\varphi_2\neq0$ for all non-trivial spinor $\psi\in\sec
{P}_{\mathrm{Spin}_{1,6}^{e}}(M)\times_{\rho}\mathbb{C}^{8}$ under the condition
(\ref{condicao}).

Next let us display the 3-form bilinear covariant:
\beq
\varphi_3 &=& \bar\psi\gamma_{k_1k_2k_3}\psi\;e^{k_1k_2k_3}\nonumber\\
&=& 2i\left(\left| \alpha_1\right| ^2-\left| \alpha_2\right| ^2+\left|
\alpha_3\right| ^2-\left| \alpha_4\right|
   ^2\right)\;e^{012}-2\left(\alpha_2\overline\alpha_1 -
\alpha_1\overline\alpha_2+\alpha_4\overline\alpha_3-\alpha_3\overline\alpha_4\right)\;e^{013}\nonumber\\
   &&-2i\left(\alpha_2\overline\alpha_1 -
\alpha_1\overline\alpha_2+\alpha_4\overline\alpha_3-\alpha_3\overline\alpha_4\right)\;e^{014}
+2\left(\alpha_2\overline\alpha_1 -
\alpha_1\overline\alpha_2+\alpha_4\overline\alpha_3-\alpha_3\overline\alpha_4\right)\;e^{016}\nonumber\\
&&+2i\left(\alpha_2\overline\alpha_1 -
\alpha_1\overline\alpha_2+\alpha_4\overline\alpha_3-\alpha_3\overline\alpha_4\right)\;e^{023}
-2\left(\alpha_2\overline\alpha_1 -
\alpha_1\overline\alpha_2+\alpha_4\overline\alpha_3-\alpha_3\overline\alpha_4\right)\;e^{024}\nonumber\\
&&-2i\left(\alpha_2\overline\alpha_1 -
\alpha_1\overline\alpha_2+\alpha_4\overline\alpha_3-\alpha_3\overline\alpha_4\right)\;e^{026}
2i\left(\left| \alpha_1\right| ^2-\left| \alpha_2\right| ^2+\left|
\alpha_3\right| ^2-\left| \alpha_4\right|
   ^2\right)\;e^{034}\nonumber\\
&&-2\left(\left| \alpha_1\right| ^2+\left| \alpha_2\right| ^2+\left|
\alpha_3\right| ^2+\left| \alpha_4\right|
   ^2\right)\;e^{036}
-2i\left(\left| \alpha_1\right| ^2-\left| \alpha_2\right| ^2+\left|
\alpha_3\right| ^2-\left| \alpha_4\right|
   ^2\right)\;e^{056}\nonumber\\
   &&+2i\left(\left| \alpha_1\right| ^2-\left| \alpha_2\right| ^2+\left|
\alpha_3\right| ^2-\left| \alpha_4\right|
   ^2\right)\;e^{126}-2i\left(\left| \alpha_1\right| ^2+\left| \alpha_2\right|
^2+\left| \alpha_3\right| ^2+\left| \alpha_4\right|
   ^2\right)\;e^{245}\nonumber\\
   &&+2i\left(\left| \alpha_1\right| ^2+\left| \alpha_2\right| ^2+\left|
\alpha_3\right| ^2+\left| \alpha_4\right|
   ^2\right)\;e^{256}-2i\left(\left| \alpha_1\right| ^2+\left| \alpha_2\right|
^2+\left| \alpha_3\right| ^2+\left| \alpha_4\right|
   ^2\right)\;e^{345}\nonumber\\
   &&\!+\!2\left(\left| \alpha_1\right| ^2\!+\!\left| \alpha_2\right|
^2\!+\!\left| \alpha_3\right| ^2\!+\!\left| \alpha_4\right|
   ^2\right)\;e^{356}\!+\!2i\left(\left| \alpha_1\right| ^2\!+\!\left|
\alpha_2\right| ^2\!+\!\left| \alpha_3\right| ^2\!+\!\left| \alpha_4\right|
   ^2\right)\;e^{456}
\eeq\noindent All terms of $\varphi_3$ have to be zero in order to $\varphi_3$
to be zero. Indeed, in particular the coefficient $\left(\left| \alpha_1\right|
^2\!+\!\left| \alpha_2\right| ^2\!+\!\left| \alpha_3\right| ^2\!+\!\left|
\alpha_4\right|
   ^2\right)$ of $e^{356}$ must be zero in order to $\varphi_4 = 0$, which
implies that $\alpha_\mu=0$.
 Hence $\varphi_3\neq0$ for all non-trivial spinor $\psi\in\sec
{P}_{\mathrm{Spin}_{1,6}^{e}}(M)\times_{\rho}\mathbb{C}^{8}$ under the condition
(\ref{condicao}).

The 4-form bilinear covariant is

\beq
\varphi_4 &=& \bar\psi\gamma_{k_1k_2k_3k_4}\psi\;e^{k_1k_2k_3k_4}\nonumber\\
&=& 2\left(\left| \alpha_1\right| ^2+\left| \alpha_2\right| ^2+\left|
\alpha_3\right| ^2+\left| \alpha_4\right|
   ^2\right)\;e^{0235}+2\left(\alpha_2\overline\alpha_1 +
\alpha_1\overline\alpha_2+\alpha_4\overline\alpha_3+\alpha_3\overline\alpha_4\right)\;e^{0345}\nonumber\\
   &&+2\left(\alpha_2\overline\alpha_1 +
\alpha_1\overline\alpha_2+\alpha_4\overline\alpha_3+\alpha_3\overline\alpha_4\right)\;e^{0356}+2\left(\alpha_2\overline\alpha_1 +
\alpha_1\overline\alpha_2+\alpha_4\overline\alpha_3+\alpha_3\overline\alpha_4\right)\;e^{0456}\nonumber\\
&&-2\left(\alpha_2\overline\alpha_1 +
\alpha_1\overline\alpha_2+\alpha_4\overline\alpha_3+\alpha_3\overline\alpha_4\right)\;e^{1236}-2i\left(\alpha_2\overline\alpha_1 -
\alpha_1\overline\alpha_2+\alpha_4\overline\alpha_3-\alpha_3\overline\alpha_4\right)\;e^{1246}\nonumber\\
&&2\left(\left| \alpha_1\right| ^2-\left| \alpha_2\right| ^2\!+\!\left|
\alpha_3\right| ^2-\left| \alpha_4\right|
   ^2\right)\;e^{1346}\!+\!2\left(\left| \alpha_1\right| ^2\!+\!\left|
\alpha_2\right| ^2\!+\!\left| \alpha_3\right| ^2\!+\!\left| \alpha_4\right|
   ^2\right)\;e^{2346}
\eeq
Again, the coefficient $\left(\left| \alpha_1\right| ^2\!+\!\left|
\alpha_2\right| ^2\!+\!\left| \alpha_3\right| ^2\!+\!\left| \alpha_4\right|
   ^2\right)$ of $e^{2346}$ must be zero in order for $\varphi_5 = 0$, implying
that $\alpha_\mu=0$.
 Hence $\varphi_4\neq0$ for all non-trivial spinor $\psi\in\sec
{P}_{\mathrm{Spin}_{1,6}^{e}}(M)\times_{\rho}\mathbb{C}^{8}$ under the condition
(\ref{condicao}).

The 5-form one is

\beq
\varphi_5&=&\bar\psi\gamma_{k_1k_2\ldots k_5}\gamma_8\,e^{k_1k_2\ldots
k_5}\;\nonumber\\&=&2\left(\left| \alpha_1\right| ^2-\left| \alpha_2\right|
^2\!+\!\left| \alpha_3\right| ^2-\left| \alpha_4\right|
   ^2\right)\;e^{01234}+2i+2\left(\alpha_2\overline\alpha_1 -
\alpha_1\overline\alpha_2+\alpha_4\overline\alpha_3-\alpha_3\overline\alpha_4\right)\;e^{02346}
\nonumber\\
&&-2\left(\left| \alpha_1\right| ^2-\left| \alpha_2\right| ^2\!+\!\left|
\alpha_3\right| ^2-\left| \alpha_4\right|
   ^2\right)\;e^{02346}
\nonumber\\
&&
+2\left(\left| \alpha_1\right| ^2+\left| \alpha_2\right| ^2\!+\!\left|
\alpha_3\right| ^2+\left| \alpha_4\right|
   ^2\right)\;e^{12345}+2i\left(\left| \alpha_1\right| ^2-\left| \alpha_2\right|
^2\!+\!\left| \alpha_3\right| ^2-\left| \alpha_4\right|
   ^2\right)\;e^{12356}\nonumber\\
&&+2\left(\alpha_2\overline\alpha_1 +
\alpha_1\overline\alpha_2+\alpha_4\overline\alpha_3+\alpha_3\overline\alpha_4\right)\;e^{23456}
\eeq
With respect to the above equation, we want to analyze in which cases we have
$\varphi_5=0$. Let us then take, in particular,  the coefficient of $e^{12345}$
to be zero, which implies that $\alpha_\mu = 0$. Hence there is no non-trivial
spinor leading to such situation.

A similar reasoning can be applied to the 6-form below:
\beq
\varphi_6&=&\bar\psi\gamma_{k_1}\gamma_8\,e^{k_1\ldots
k_6}\;\nonumber\\&=&-2i\left(\alpha_2\overline\alpha_1 -
\alpha_1\overline\alpha_2+\alpha_4\overline\alpha_3-\alpha_3\overline\alpha_4\right)\;e^{012346}-2\left(\left| \alpha_1\right| ^2+\left| \alpha_2\right|
^2\!-\!\left| \alpha_3\right| ^2-\left| \alpha_4\right|
   ^2\right)\;e^{123456}\nonumber\\
  &&-2\left(\left| \alpha_1\right| ^2\!+\!\left| \alpha_2\right| ^2\!-\!\left|
\alpha_3\right| ^2\!-\!\left| \alpha_4\right|
   ^2\right)e^{013456}+2i\left(\alpha_2\overline\alpha_1 \!-\!
\alpha_1\overline\alpha_2\!+\!\alpha_4\overline\alpha_3\!-\!\alpha_3\overline\alpha_4\right)\;e^{012456}\label{6-form}
\eeq

Finally we calculate the 7-form:
\beq
\varphi_7&=&2i\left(\alpha_2\overline\alpha_1 -
\alpha_1\overline\alpha_2+\alpha_4\overline\alpha_3-\alpha_3\overline\alpha_4\right)\;e^{012456}\label{7-form}
\eeq


\footnotesize
\begin{thebibliography}{99}
\footnotesize

 \bibitem{Cavalcanti:2014wia}
  R.~T.~Cavalcanti,
  \emph{Looking for the Classification of Singular Spinor Fields Dynamics and
other Mass Dimension One Fermions: Characterization of Spinor Fields}, Int. J.
Mod. Phys. D {\bf 23} (2014) 1444002 
  [{\tt arXiv:1408.0720 [hep-th]}].  


\bibitem{daSilva:2012wp}
  J.~M.~Hoff da Silva and R.~da Rocha,
 \emph{Unfolding Physics from the Algebraic Classification of Spinor Fields,}
  Phys.\ Lett.\ B {\bf 718} (2013) 1519
  [{\tt arXiv:1212.2406 [hep-th]}].










\bibitem{exotic} R.~da Rocha, A.~E.~Bernardini and J.~M.~Hoff da Silva,
\emph{Exotic Dark Spinor Fields}, \emph{JHEP} {\bf 04} (2011) 110 [{\tt
arXiv:1103.4759 [hep-th]}].

\bibitem{lee2}
  D.~V.~Ahluwalia, C.~-Y.~Lee, D.~Schritt and T.~F.~Watson,
  \emph{Elko  as self-interacting fermionic dark matter with axis of locality}, 
 Phys.\ Lett.\ B {\bf 687} (2010) 248  [{\tt arXiv:0804.1854 [hep-th]}].


\bibitem{esk}
  R.~da Rocha, L.~Fabbri, J.~M.~Hoff da Silva, R.~T.~Cavalcanti and
J.~A.~Silva-Neto, \emph{Flag-Dipole Spinor Fields in ESK Gravities}, J.\ Math.\
Phys.\  {\bf 54} (2013) 102505
  [{\tt arXiv:1302.2262 [gr-qc]}].

\bibitem{daRocha:2007pz}
  R.~da Rocha and J.~M.~Hoff da Silva,
  \emph{From Dirac spinor fields to ELKO}, 
  J.\ Math.\ Phys.\  {\bf 48} (2007) 123517 [{\tt arXiv:0711.1103 [math-ph]}].

 \bibitem{bht} R.~da Rocha and J.~M.~Hoff da Silva,
  \emph{Hawking Radiation from Elko  Particles Tunnelling across Black Strings
Horizon,}
  Europhys.\ Lett.\  {\bf 107} (2014) 50001
  [{\tt arXiv:1408.2402 [hep-th]}].


\bibitem{m1}
  M.~Dias, F.~de Campos and J.~M.~Hoff da Silva,
  \emph{Exploring Elko  typical signature,} 
  Phys.\ Lett.\ B {\bf 706} (2012) 352 
  [{\tt arXiv:1012.4642}].



\bibitem{bonora}
  L.~Bonora, K.~P.~S.  Brito and R.~da Rocha,
  \emph{Spinor Fields Classification in Arbitrary Dimensions and New Classes of
Spinor Fields on 7-Manifolds,} 
  JHEP {\bf 1502} (2015) 069
  [{\tt arXiv:1411.1590 [hep-th]}].

\bibitem{1} C. I. Lazaroiu, E. M. Babalic, I. A. Comam, \emph{The geometric
algebra of Fierz identities in arbitrary dimensions and signatures}, {JHEP}
{\bf 09} (2013) 156 [{\tt arXiv:1304.4403 [hep-th]}].



\bibitem{SUGRA} E. Cremmer, B. Julia, J. Scherk, \emph{Supergravity Theory in
11 Dimensions}, Phys. Lett. B {\bf 76} (1978) 409.

\bibitem{SUSYBr} F. Englert, M. Rooman, P. Spindel, \emph{Supersymmetry Breaking
by Torsion and The Ricci-Flat Squashed Seven-Spheres}, Phys. Lett. B {\bf 127}
(1983) 47.

\bibitem{Duff:1986hr}
  M.~J.~Duff, B.~E.~W.~Nilsson and C.~N.~Pope,
 \emph{Kaluza-Klein Supergravity}, 
  Phys.\ Rept.\  {\bf 130} (1986) 1.

\bibitem{mei10}  J.~Mei,
{\it Entropy for General Extremal Black Holes,}  JHEP {\bf
04}  (2010) 005 [{\tt arXiv:1002.1349 [hep-th]}].

\bibitem{Ahluwalia:2008xi}
  D.~V.~Ahluwalia, C.~Y.~Lee, D.~Schritt and T.~F.~Watson,
  \emph{Elko as self-interacting fermionic dark matter with axis of locality}, 
  Phys.\ Lett.\ B {\bf 687} (2010) 248
  [{\tt arXiv:0804.1854 [hep-th]}].

\bibitem{Ahluwalia:2004ab}
  D.~V.~Ahluwalia and D.~Grumiller,
  \emph{Spin half fermions with mass dimension one: Theory, phenomenology, and dark matter}, 
  JCAP {\bf 0507} (2005) 012
  [{\tt hep-th/0412080}].

 \bibitem{Cra} J. P. Crawford, \emph{On The Algebra Of Dirac Bispinor Densities:
Factorization And Inversion Theorems},  J. Math. Phys. \textbf{26} (1985) 1439.


\bibitem{Taka} Y. Takahashi, \emph{Reconstruction of Spinor From Fierz
Identities}, Phys. Rev. D \textbf{26} (1982) 2169.

 

 \bibitem{wei}
  J.~Mei,
  \emph{Spinor Fields and Symmetries of the Spacetime}, 
  Gen.\ Rel.\ Grav.\  {\bf 44} (2012) 2191
  [{\tt arXiv:1105.5741 [gr-qc]}].


 \bibitem{budinich}
  P. Budinich,
  \emph{From the geometry of pure spinors with their division algebras to
fermion's physics},
  Found.\ Phys.\  {\bf 32} (2002) 1347
  [{\tt hep-th/0107158}].
 
 \bibitem{Babalic:2014cea}
  E.~M.~Babalic and C.~I.~Lazaroiu,
  \emph{A generalization of Calabi-Yau fourfolds arising from M-theory
compactifications}, 
  Bulg.\ J.\ Phys.\  {\bf 41} (2014) 109
  [{\tt arXiv:1411.3493 [hep-th]}].



\bibitem{alex}
 A.~E.~Bernardini and R.~da Rocha,
  \emph{Dynamical dispersion relation for Elko   dark spinor fields},
 Phys.\ Lett.\ B {\bf 717} (2012) 238
  [{\tt arXiv:1203.1049 [hep-th]}].
\end{thebibliography}
\end{document}